\documentclass{appolb}
\usepackage{graphicx}
\newcommand{\delhad}{\mbox{$\Delta \alpha_{\rm had}^{(5)}(M_Z)$}} 
                   
\newcommand{\msb}{\mbox{$\overline{\rm{MS}}\ $}}                                
\newcommand{\mt}{\mbox{$m_t$}}                                                  
\newcommand{\mh}{\mbox{$M_H$}}                                                  
\newcommand{\mz}{\mbox{$M_Z$}}                                                  
\newcommand{\mw}{\mbox{$M_W$}}                                                  
                                      
\newcommand{\als}{\mbox{$\alpha_s$}}                                            
\newcommand{\suf}{\mbox{$SU(5)\ $}}

\newcommand{\skipblk}[1]{}                                                      
\def\bqa{\begin{eqnarray}}                                                      
\def\eqa{\end{eqnarray}}                                                        
\newcommand{\ee}{\mbox{$e^+ e^-$}}

\newcommand{\sto}{\mbox{$SU(2) \x U(1)\ $}}                                       
                                               
\newcommand{\x}{\mbox{$\times$}}

\newcommand{\sinn}{\mbox{$\sin^2\theta_W\,$}}                                   
                                            
\newcommand{\snu}{\mbox{$\stackrel{(-)}{\nu}$}}                                 
\newcommand{\beq}{\begin{equation}}                                             
\newcommand{\eeq}{\end{equation}}

\newcommand{\RA}{\mbox{$\rightarrow$}}

\def\mxth{\mathsurround=0pt }
\def\xversim#1#2{\lower2.pt\vbox{\baselineskip0pt \lineskip-.5pt
  \ialign{$\mxth#1\hfil##\hfil$\crcr#2\crcr\sim\crcr}}}             
\def\simgr{\mathrel{\mathpalette\xversim >}}                                    
\def\simle{\mathrel{\mathpalette\xversim <}}                                    

\begin{document}
\title{Electroweak Physics
\thanks{Presented at XXXVI International Meeting on Fundamental Physics, 
Baeza, Spain, February 2008.}}%
\author{Jens Erler
\address{Departamento de F\'isica Te\'orica, Instituto de F\'isica, 
Universidad Nacional Aut\'onoma de M\'exico, M\'exico D.F. 04510, M\'exico}
\and
Paul Langacker
\address{Institute for Advanced Study, Princeton, NJ  08540, U.S.A.}}
%\and
%the Name(s) of other Author(s)
%\address{and their affiliation}
\maketitle
\begin{abstract}
The results of high precision weak neutral current
(WNC), $Z$-pole, and high energy collider electroweak experiments 
have been the primary prediction and test of electroweak unification.
The electroweak program is briefly reviewed from a historical perspective.
The current status 
and the implications for the standard model and beyond are discussed.
\end{abstract}
  
\section{The $Z$, the $W$, and the Weak Neutral Current}

The weak neutral current was a critical prediction of the electroweak
standard model (SM)~\cite{sirlinfest,general}. Following its discovery in 1973
by the Gargamelle and HPW experiments, there
were generations of ever more precise WNC experiments, typically at the
few~\% level.  These included
pure weak $\nu N$ and $\nu e$ scattering processes, and
weak-electromagnetic interference processes such as polarized
$e^{\uparrow \downarrow}D$ or $\mu N$, $\ee \RA $ (hadron or charged lepton)
cross sections and asymmetries below the $Z$ pole, and parity-violating
effects in heavy atoms (APV).  There were also  early direct observations
of the $W$ and $Z$ by UA1 and UA2. The early 1990's witnessed the very precise
$Z$-pole experiments at LEP and the SLC, in which the lineshape, decay modes,
and various asymmetries were measured at the 0.1\% level. The subsequent
LEP 2 program at higher energies measured \mw, searched for the Higgs
and other new particles, and constrained anomalous gauge self-interactions.
Parallel efforts at the Tevatron by CDF and D\O \
led to the direct discovery of the $t$ and  measurements of \mt \ and \mw,
while a fourth generation of weak neutral current experiments
continued to search for new physics to which the (more precise) $Z$-pole
experiments were blind. 
The program
was supported by theoretical efforts in the
calculation of QCD, electroweak, and mixed
radiative corrections; the expectations for observables
in the standard model,
large classes of extensions, and alternative models; and global
analyses of the data.

The precision program
has established
that the standard model (SM) is correct and unique to first approximation,
establishing the gauge
principle as well as the SM gauge group and representations;
shown that the SM is correct at loop level, confirming the basic principles of
renormalizable gauge theory and allowing the successful prediction 
or constraint on $m_t$, $\alpha_s$, and the Higgs mass $M_H$;
severely constrained new  physics at the TeV scale, with the
ideas of unification favored over TeV-scale dynamics or   compositeness; and
yielded precise values for the  gauge couplings, consistent with 
(supersymmetric) gauge unification.

\section{Results before the LEP/SLD era}

Even before the beginning of the $Z$-pole experiments at LEP and SLC in 1989,
the precision  program had established~\cite{general}-\cite{costa}:
\begin{itemize}
\item Global analyses of all data carried more information than
the analysis of individual experiments, but care has to be taken
with systematic and theoretical uncertainties.
\item The SM is correct to first approximation.
The four-fermion operators for $\nu q$, $\nu e$,
and $eq$ were uniquely determined,
in agreement with the standard model, in model (i.e., gauge group) independent 
analyses. 
The $W$ and $Z$ masses agreed with the expectations
of the \sto gauge group and canonical Higgs mechanism, eliminating more complicated
alternative models with the
same four-fermi interactions as the standard model.
       \item QCD evolved structure functions and 
      electroweak radiative corrections  were necessary  for the agreement
of theory and experiment.
        \item The weak mixing angle (in the on-shell renormalization scheme) 
was determined to be \sinn = 0.230 $\pm 0.007$; consistency of the various
observations, including radiative corrections,  required
$m_t < 200$ GeV.
\item Theoretical uncertainties, especially in the $c$ threshold 
in deep inelastic weak  charge current (WCC) scattering,
dominated.
        \item The combination of WNC and WCC data uniquely
determined the $SU(2)$ representations of all of the known fermions,
i.e.,  $\nu_e$ and $\nu_\mu$, as well as the $L$ and
$R$ components of the $e, \ \mu, \ \tau, \ d, \, s, \, b, \ u,$ and $c$~\cite{unique}.
In particular,  the left-handed $b$ and $\tau$ were  the
lower components of $SU(2)$ doublets, implying unambiguously that the $t$ quark
and $\nu_\tau$ had to exist.
This was independent of theoretical arguments based on
anomaly cancellation (which could have been evaded in alternative models
involving a vector-like third family), and of
constraints on \mt \ from electroweak loops.
        \item The electroweak gauge couplings were
well-determined, allowing a detailed comparison with the gauge
unification predictions of the simplest grand unified theories (GUT).
Ordinary
    \suf was excluded (consistent with the non-observation of proton decay),
but the supersymmetric extension was allowed, ``perhaps even the first harbinger of supersymmetry''~\cite{amaldi}.
        \item There were stringent limits on new physics at the TeV scale, including
additional $Z'$ bosons, exotic fermions (for which both WNC and WCC 
constraints were crucial), exotic Higgs representations,
 leptoquarks, and new four-fermion operators.
\end{itemize}

\section{The LEP/SLC Era}
The LEP/SLC era greatly improved the precision of the electroweak program.
It allowed the differentiation between non-decoupling extensions to the
SM (such as most forms of dynamical symmetry breaking and other types
of TeV-scale compositeness), which typically predicted several
\% deviations, and decoupling extensions (such as most of the 
parameter space for supersymmetry), for which the deviations are
typically 0.1\%.

The first phase of the LEP/SLC program involved running at the $Z$
pole, $e^+ e^- \rightarrow Z \rightarrow \ell^+ \ell^-, \ \
  q \bar{q},$ and $\nu \bar{\nu}$. During the period 1989-1995 the
four LEP experiments ALEPH, DELPHI, L3, and OPAL at CERN
observed $\sim  2 \times 10^{7} Z$ bosons. The SLD experiment at the SLC at
SLAC observed some $5 \times 10^5$ events. Despite the much lower statistics,
the SLC had the considerable advantage of a highly polarized $e^-$ beam,
with $P_{e^-} \sim$ 75\%. There were quite a few $Z$ pole observables,
including:
 \begin{itemize}
   \item The lineshape: $M_Z, \Gamma_Z,$ and the peak cross section $ \sigma$.
   \item The branching ratios for $e^+e^-,\ \mu^+ \mu^-,\ \tau^+ \tau^-, 
\ q \bar{q},\ c \bar{c},\ b \bar{b},$ and $ s \bar{s}$. One could also determine
the invisible width, $\Gamma({\rm inv})$, from which
one can derive the number 
$N_\nu = 2.985
\pm 0.009$ of active (weak doublet) neutrinos with 
       $m_\nu < M_Z/2$, i.e., there are only 3 conventional families with 
light neutrinos. $\Gamma({\rm inv})$ also constrains other invisible
particles, such as light sneutrinos and the light majorons associated with some 
models of neutrino mass.
   \item A number of asymmetries, including forward-backward (FB) asymmetries; 
the $\tau$ polarization, $P_\tau$;  the polarization asymmetry $A_{LR}$ associated
with $P_{e^-}$; and
mixed polarization-FB asymmetries.
    \end{itemize}
The expressions for the observables are summarized in~\cite{sirlinfest,general},
and the experimental values and SM predictions in
Table~\ref{tab1}.
The precision of the $Z$ mass determination was extraordinary for a high
energy experiment.
These combinations of observables could be used to isolate many
$Z$-fermion couplings, verify lepton family universality,
determine \sinn in numerous ways, and determine or constrain \mt, \als, and \mh.
 LEP and SLC simultaneously carried out other  programs,
most notably studies and tests of QCD, and heavy quark physics.

%:Table 1

\small

\begin{table} \centering
\begin{tabular}{|l|c|c|c|r|}\hline
 Quantity & {\rm Value} & {\rm Standard Model} & {\rm Pull} & {\rm Dev.}                       \cr %\noalign{\vskip -.1in}
\hline
$M_Z$                    [GeV]   &  $91.1876 \pm 0.0021$ &  $91.1874 \pm 0.0021$  & $ 0.1$     &$ -0.1  $   \cr%\noalign{\vskip 2pt}
$\Gamma_Z$               [GeV]   & $   2.4952 \pm 0.0023 $&  $   2.4968 \pm 0.0010  $&  $ -0.7     $&  $ -0.5     $ \cr%\noalign{\vskip 2pt}
$\Gamma({\rm had})$      [GeV]   & $   1.7444 \pm 0.0020 $&  $   1.7434 \pm 0.0010  $&  \hbox{---}&\hbox{---}\cr%\noalign{\vskip2pt}  
$\Gamma({\rm inv})$      [MeV]   & $ 499.0    \pm 1.5    $&  $ 501.59   \pm 0.08    $& \hbox{---}&\hbox{---}\cr%\noalign{\vskip 2pt}
$\Gamma({\ell^+\ell^-})$ [MeV]   & $  83.984  \pm 0.086  $&  $  83.988  \pm 0.016   $& \hbox{---}&\hbox{---}\cr%\noalign{\vskip 2pt} 
$\sigma_{\rm had}$       [nb]    & $  41.541  \pm 0.037  $&  $  41.466  \pm 0.009   $&  $  2.0     $&  $  2.0     $ \cr%\noalign{\vskip 2pt}
$R_e$                            & $  20.804  \pm 0.050  $&  $  20.758  \pm 0.011   $&  $  0.9     $&  $  1.0     $ \cr%\noalign{\vskip 2pt}
$R_\mu$                          & $  20.785  \pm 0.033  $&  $  20.758  \pm 0.011   $&  $  0.8     $&  $  0.9     $ \cr%\noalign{\vskip 2pt}
$R_\tau$                         & $  20.764  \pm 0.045  $&  $  20.803  \pm 0.011   $&  $ -0.9     $&  $ -0.8     $ \cr%\noalign{\vskip 2pt}
$R_b$                            & $   0.21629\pm 0.00066$&  $   0.21584\pm 0.00006 $&  $  0.7     $&  $  0.7     $ \cr%\noalign{\vskip 2pt}
$R_c$                            & $   0.1721 \pm 0.0030 $&  $   0.17228\pm 0.00004 $&  $ -0.1     $&  $ -0.1     $ \cr%\noalign{\vskip 2pt}
$A_{FB}^{(0,e)}$                 & $   0.0145 \pm 0.0025 $&  $   0.01627\pm 0.00023 $&  $ -0.7     $&  $ -0.6     $ \cr%\noalign{\vskip 2pt}
$A_{FB}^{(0,\mu)}$               & $   0.0169 \pm 0.0013 $&                        & $  0.5     $&  $  0.7     $ \cr%\noalign{\vskip 2pt}
$A_{FB}^{(0,\tau)}$              &  $   0.0188 \pm 0.0017 $&                       & $  1.5     $&  $  1.6     $ \cr%\noalign{\vskip 2pt}
$A_{FB}^{(0, b  )}$              & $   0.0992 \pm 0.0016 $&  $   0.1033 \pm 0.0007  $&  $ -2.5     $&  $ -2.0     $ \cr%\noalign{\vskip 2pt}
$A_{FB}^{(0, c  )}$              & $   0.0707 \pm 0.0035 $&  $   0.0738 \pm 0.0006  $&  $ -0.9     $&  $ -0.7     $ \cr%\noalign{\vskip 2pt}
$A_{FB}^{(0, s  )}$              & $   0.0976 \pm 0.0114 $&  $   0.1034 \pm 0.0007  $&  $ -0.5     $&  $ -0.4     $ \cr%\noalign{\vskip 2pt}
$\bar{s}_\ell^2(A_{FB}^{(0,q)})$ & $   0.2324 \pm 0.0012 $&  $   0.23149\pm 0.00013 $&  $  0.8     $&  $  0.6     $ \cr%\noalign{\vskip 2pt}
                                 & $   0.2238 \pm 0.0050 $&                        & $ -1.5     $&  $ -1.6     $ \cr%\noalign{\vskip 2pt}
$A_e$                            & $   0.15138\pm 0.00216$&  $   0.1473 \pm 0.0011  $&  $  1.9     $&  $  2.4     $ \cr%\noalign{\vskip-4pt}
                                 & $   0.1544 \pm 0.0060 $&                        & $  1.2     $&  $  1.4     $ \cr%\noalign{\vskip-4pt}
                                & $   0.1498 \pm 0.0049 $&                       & $  0.5     $&  $  0.7     $ \cr%\noalign{\vskip 2pt}
$A_\mu$                          & $   0.142  \pm 0.015  $&                       & $ -0.4     $&  $ -0.3     $ \cr%\noalign{\vskip 2pt}
$A_\tau$                         & $   0.136  \pm 0.015  $&                       & $ -0.8     $&  $ -0.7     $ \cr%\noalign{\vskip-4pt}
                                 & $   0.1439 \pm 0.0043 $&                        & $ -0.8     $&  $ -0.5     $ \cr%\noalign{\vskip 2pt}
$A_b$                            & $   0.923  \pm 0.020  $&  $   0.9348 \pm 0.0001  $&  $ -0.6     $&  $ -0.6     $ \cr%\noalign{\vskip 2pt}
$A_c$                            & $   0.670  \pm 0.027  $&  $   0.6679 \pm 0.0005  $&  $  0.1     $&  $  0.1     $ \cr%\noalign{\vskip 2pt}
$A_s$                            & $   0.895  \pm 0.091  $&  $   0.9357 \pm 0.0001  $&  $ -0.4     $&  $ -0.4     $ \cr%\noalign{\vskip 2pt}
\hline
\end{tabular}
\caption{Principal $Z$-pole observables, their experimental values, 
theoretical predictions using the SM parameters from the global best
fit with $M_H$ free (yielding $M_H=77^{+28}_{-22}$ GeV), pull
(difference from the prediction divided by the uncertainty), and Dev. (difference for fit
with $M_H$ fixed at 117 GeV, just above the direct search limit of 114.4 GeV), as of 11/07, from~\cite{general}.
See~\cite{sirlinfest,general} for definitions of the quantitites.
$\Gamma({\rm had})$, $\Gamma({\rm inv})$, and $\Gamma({\ell^+\ell^-})$ are not 
independent.}
\label{tab1}
\end{table}

\normalsize

LEP~2 ran from 1995-2000, with energies gradually increasing from $\sim 140$ to $\sim 209$ GeV.
The principal electroweak results were precise measurements of the $W$ mass, as well
as its width and branching ratios (these were measured independently at the Tevatron);
a measurement of  $e^+ e^- \RA W^+ W^-$,  $ZZ$, and single $W$,
as a function of center of mass (CM)
energy, which tests the cancellations between diagrams that is characteristic
of a renormalizable gauge field theory, or, equivalently, probes the triple
gauge vertices;
limits on anomalous quartic gauge vertices;
measurements of various cross sections and asymmetries for
$e^+ e^- \RA f \bar{f}$ for $f=\mu^-,\tau^-,q,b$ and $c$, in reasonable
agreement with SM predictions;
a stringent lower limit of 114.4 GeV on the Higgs mass, and even hints
of an observation at $\sim$ 116 GeV;
and searches for supersymmetric or other exotic particles.

In parallel with the LEP/SLC program, there were 
precise ($< $ 1\%) measurements of atomic parity violation (APV) in cesium at Boulder,
along with the atomic calculations and related measurements needed for the
interpretation; precise new measurements of deep inelastic
scattering by the NuTeV collaboration at Fermilab, with
a sign-selected beam which allowed them to minimize the effects of the $c$ threshold
and reduce uncertainties to around 1\%; and few \% measurements of $\snu_\mu e$ by CHARM II
at CERN. Although the precision of these WNC processes was  lower
than the $Z$ pole measurements, they are still of considerable importance:
the $Z$ pole  experiments are blind to  types of new physics
that do not directly affect the $Z$,
such as a heavy $Z'$ if there is no $Z-Z'$ mixing,  while the WNC experiments are often very
sensitive. During the same period there were important electroweak results 
from CDF and D$\not{\! 0}$ at the Tevatron, most notably a precise value for $M_W$,
competitive with and complementary to the LEP~2 value; a direct measure of \mt,
and direct searches for  $Z'$, $W'$, exotic fermions, and supersymmetric particles.
The Tevatron program continues to the present day, and there have been recent precise measurements
of the  $e^-e^-$ (M\o ller) polarization asymmetry at SLAC; polarization asymmetries in $e^-$-hadron scattering at
MIT-Bates, Mainz, and Jefferson Lab; asymmetry measurements at the Tevatron; and measurements of
the $W$ propagator and of $Z$ exchange effects at HERA.
Many of these non-$Z$ pole results are summarized in
Table~\ref{tab2}.

%:Table tab2

\small

\begin{table} \centering
\begin{tabular}{|l|c|c|c|r|}\hline
 Quantity & {\rm Value} & {\rm Standard Model} & {\rm Pull} & {\rm Dev.}                       \cr %\noalign{\vskip -.1in}
\hline
$m_t$                    [GeV]  &$ 170.9\pm 1.8\pm 0.6 $& $171.1    \pm 1.9$     &$ -0.1$     &$ -0.8  $   \cr%\noalign{\vskip 2pt}
%$M_W$                    [GeV] ($\bar p p$)   & $ 80.428  \pm 0.039$  & $ 80.375  \pm 0.015 $  &  1.4     &  1.7     \cr%\noalign{\vskip-4pt}
$M_W$   ($\bar p p$)   & $ 80.428  \pm 0.039$  & $ 80.375  \pm 0.015 $  &  1.4     &  1.7     \cr%
$M_W$ (LEP)                                & $ 80.376  \pm 0.033 $ &                      &  0.0     &  0.5     \cr%\noalign{\vskip 2pt}
$g_L^2$                          &  $ 0.3010 \pm 0.0015$ &$   0.30386\pm 0.00018 $&$ -1.9    $ &$ -1.8 $    \cr%\noalign{\vskip 2pt}
$g_R^2$                          &   $0.0308 \pm 0.0011$ & $  0.03001\pm 0.00003$ &  0.7     &  0.7     \cr%\noalign{\vskip 2pt}
$g_V^{\nu e}$                    & $ -0.040  \pm 0.015 $ & $ -0.0397 \pm 0.0003 $ &  0.0     &  0.0     \cr%\noalign{\vskip 2pt}
$g_A^{\nu e}$                    & $ -0.507  \pm 0.014 $ & $ -0.5064 \pm 0.0001 $ &  0.0     &  0.0     \cr%\noalign{\vskip 2pt} &
$A_{PV} \times 10^7$ &$ -1.31 \pm 0.17$ &$ -1.54 \pm 0.02 $ &  1.3     &  1.2     \cr%\noalign{\vskip 2pt} 
$Q_W({\rm Cs})$                  &$ -72.62   \pm 0.46 $  &$ -73.16   \pm 0.03$    &  1.2     &  1.2     \cr%\noalign{\vskip 2pt} 
$Q_W({\rm Tl})$                  &$-116.4    \pm 3.6 $   &$-116.76   \pm 0.04    $&  0.1     &  0.1     \cr%\noalign{\vskip 2pt} 
${\Gamma(b\rightarrow s\gamma)\over\Gamma(b\rightarrow X e\nu)}$ &$ \left(3.55^{+0.53}_{-0.46}\right)\times 10^{-3} $
                                              &$ (3.19 \pm 0.08) \times 10^{-3} $&  0.8     &  0.7     \cr%\noalign{\vskip 2pt}
${1\over 2}(g_\mu-2-{\alpha\over \pi})$      &  $   4511.07(74) \times 10^{-9}$
                                              &    $ 4509.08(10) \times 10^{-9}$ &  2.7     &  2.7     \cr%\noalign{\vskip 2pt} 
$\tau_\tau$ [fs]                 &$ 290.93   \pm 0.48  $ &$ 291.80   \pm 1.76 $   &$ -0.4 $    &$ -0.4$     \cr%\noalign{\vskip 2pt}
\hline
\end{tabular}
\caption{Non-$Z$-pole observables, 11/07. The SM values are  from~\cite{general}.}
\label{tab2}
\end{table}

\normalsize

The effort required
the calculation of the needed electromagnetic,
electroweak,  QCD, and mixed radiative corrections
to the predictions of the SM. Careful consideration of
the competing definitions of the renormalized \sinn
was needed. 
The principal theoretical uncertainty is the hadronic
contribution \delhad \ to the running of 
$\alpha$ from its precisely known value at low energies
to the $Z$-pole, where it is needed to compare
the $Z$ mass with the asymmetries and other observables.
The radiative corrections, renormalization schemes, and
running of $\alpha$ are further discussed in~\cite{sirlinfest,general}.
The LEP Electroweak Working Group (LEPEWWG)~\cite{LEPEWWG} 
combined the results of
the four LEP experiments, and also those of SLD and some WNC and Tevatron
results, taking proper account of
common systematic and theoretical uncertainties.
Much theoretical effort  also went into the development,
testing, and comparison of radiative corrections packages, and
into the study of how various classes of new
physics would modify the observables, and how they could
most efficiently be parametrized.

\section{Comments on the Data}

\begin{itemize}

\item As can be seen in Table \ref{tab1} most of the $Z$-pole measurements are
in excellent agreement with the standard model predictions using the
parameters from the global best fit. One exception is the LEP value for
$A_{FB}^{(0, b  )}$, the forward-backward asymmetry into $b$ quarks,
which is 2.5$\sigma$ below the best fit expectation, and 2.0$\sigma$
below the fit with $M_H=117$ GeV. If not just a statistical fluctuation or systematic problem, $A_{FB}^{(0, b  )}$ could
be a hint of new physics.  However, any such effect should not contribute
too much to  $R_b$. The deviation is only around 3.9\%, but if the
new physics involved a 
 radiative correction to the coefficient $\kappa$ \
of \sinn, the change would have to be around
20\%. Hence, the new physics  would most likely be at tree-level type affecting preferentially 
the third generation. Examples include the decay of a scalar neutrino 
resonance \cite{Erler97}, mixing of the $b$ quark with heavy 
exotics \cite{Choudhury:2001hs}, and a heavy $Z'$ with family-nonuniversal 
couplings \cite{zpr,fcnc}.

\item
There is a strong correlation between $A_{FB}^{(0, b  )}$ and
the predicted Higgs mass $M_H$ in the global fits, and
in fact a fit to $A_{FB}^{(0, b  )}, A_{FB}^{(0, c  )},$ and $M_Z$ alone yields
a prediction
$M_H=326^{+224}_{-136}$ GeV \cite{general}. In contrast, the SLD polarization asymmetry 
$A_{LR}$ combined with $M_Z$ yields a lower value $M_H=25^{+23}_{-15}$ GeV,
with the other measurements closer to the average $77^{+28}_{-22}$ GeV.
It
has been emphasized~\cite{chanowitz} that if one eliminated $A_{FB}^{(0, b  )}$ from the
fit (e.g., because it is affected by new physics) then the global fit prediction for  $M_H$
would be  lowered, with the central value well below the
lower limit of $114.4$ GeV from the direct searches at LEP 2.
One resolution, assuming $A_{FB}^{(0, b  )}$  is due to new physics or a large fluctuation, is to invoke a supersymmetric extension
of the standard model with light sparticles and second Higgs doublet~\cite{Heinemeyer:2007bw}, which modify the radiative corrections and Higgs constraints.

\item The NuTeV collaboration at Fermilab~\cite{Nutev} has
reported the  results of its deep inelastic measurements of
$\frac{\snu_\mu N \RA \snu_\mu X}{\snu_\mu N \RA \mu^{\mp}X}$.
They  greatly reduce the
uncertainty in the charm quark threshold in the charged current denominator
by taking appropriate combinations of $\nu_\mu$ and $\bar{\nu}_\mu$.
They find a value for the on-shell weak angle $s_W^2$ of
0.2277(16), which is 3.0$\sigma$ above the global fit value of 0.2231(3).
Most of the difference is in the left  handed neutral current
coupling $g_L^2$. The discrepancy is reduced to $\sim 2\sigma$ if one
incorporates (as is done here) the effects of the difference between the strange and 
antistrange quark momentum distributions, 
$S^- \equiv \int_0^1 {\rm d} x x [s(x) - \bar{s}(x)] = 0.00196 \pm 0.00135$,
from dimuon events, recently reported by NuTeV~\cite{Mason:2006fk}.
Other possible effects that could contribute are large isospin violation in
the nucleon sea,  next to leading
order QCD effects  and electroweak corrections, and nuclear shadowing~\cite{general,davidson}.
A full reanalysis of all deep inelastic data taking into account these issues and all of the uncertainties
would be extremely useful. Possible new physics explanations of the NuTeV anomaly, such as a $Z'$
with specific couplings and neutrino mixing are reviewed in~\cite{davidson}.

\item The Brookhaven $g_{\mu}-2$ experiment has reported a very precise
value~\cite{gmin2}, leading to a world average
$ a_\mu^{\rm exp} = {g_\mu - 2 \over 2} = (1165920.80 \pm 0.63) \times 10^{-9}$. 
The QED contribution has been calculated to four 
loops, and the predicted SM 
electroweak contribution is
$a_\mu^{\rm EW} = (1.52\pm 0.03) \times 10^{-9}$~\cite{gm2rev}.
The largest uncertainty in the standard model prediction  is from
the hadronic vacuum polarization contribution, which has been estimated to two loops.
This cannot be calculated perturbatively, but involves a dispersion relation
that can be evaluated using experimental data from $e^+e^-\RA$ hadrons or hadronic $\tau$ decays.
A recent analysis~\cite{davier} indicates a $3.3\sigma$ discrepancy between the standard model
prediction and the experimental value of $ a_\mu$ when the $e^+e^-$ data are used, but
only a 0.9$\sigma$ difference using the hadronic $\tau$ decays. The issue is still not settled, but
most recent authors advocate the $e^+e^-$ value because the $\tau$ decays involve uncertainties
from isospin violation. There is also a small but hard to pin down uncertaintly from the hadronic light by light scattering diagrams.

Because of the confused situation with the vacuum polarization,
it is hard to know how seriously to take the discrepancy.
Nevertheless, $a_\mu $ is more sensitive than the electron moment to most
types of new physics, so it is important.
One obvious candidate for a new physics explanation would be supersymmetry~\cite{wjm},
with relatively low masses for the relevant sparticles and high $\tan \beta$
(roughly, one requires an effective mass scale of $\tilde{m} \sim 55 \ {\rm GeV} \
\sqrt{\tan \beta}$). There is a correlation between
the theoretical uncertainty in the vacuum polarization and in the
hadronic contribution to the running of $\alpha$ to the $Z$ pole~\cite{corr},
leading to a slight reduction in the predicted Higgs mass when $a_\mu$
is included in the global fit assuming the standard model  (as is done here).

\item \delhad, the hadronic contribution to the running of $\alpha$ up to the $Z$-pole,
introduces the largest theoretical uncertainty into the precision program, in
particular to the relation between \mz \ and the \msb \  weak angle $\hat{s}^2_Z$
(extracted mainly from the asymmetries). The uncertainty is closely related to that in
$a^{\rm had}_\mu$.

\item 
The LEP and SLC $Z$-pole experiments are the most precise tests of the
standard electroweak theory, but they are insensitive to any new
physics that doesn't affect the $Z$ or its couplings. Non-$Z$-pole
experiments are therefore extremely important, especially given
the possible NuTeV anomaly. The recent measurement of 
 polarized M\o ller scattering from SLAC~\cite{moller}
 is in agreement with the standard model and observes the running of
 the weak mixing angle in the \msb scheme at the 6.4$\sigma$ level, as can be
 seen in Figure \ref{s2w}.
 An even more precise result
 is anticipated from the Qweak polarized electron experiment at Jefferson Lab~\cite{qweak}.
 The running is observed  to even lower $Q^2$
 in atomic parity violation~\cite{apv}.
%:Figure s2w
\begin{figure}[htbp]
\begin{center}
 \includegraphics*[scale=0.5]{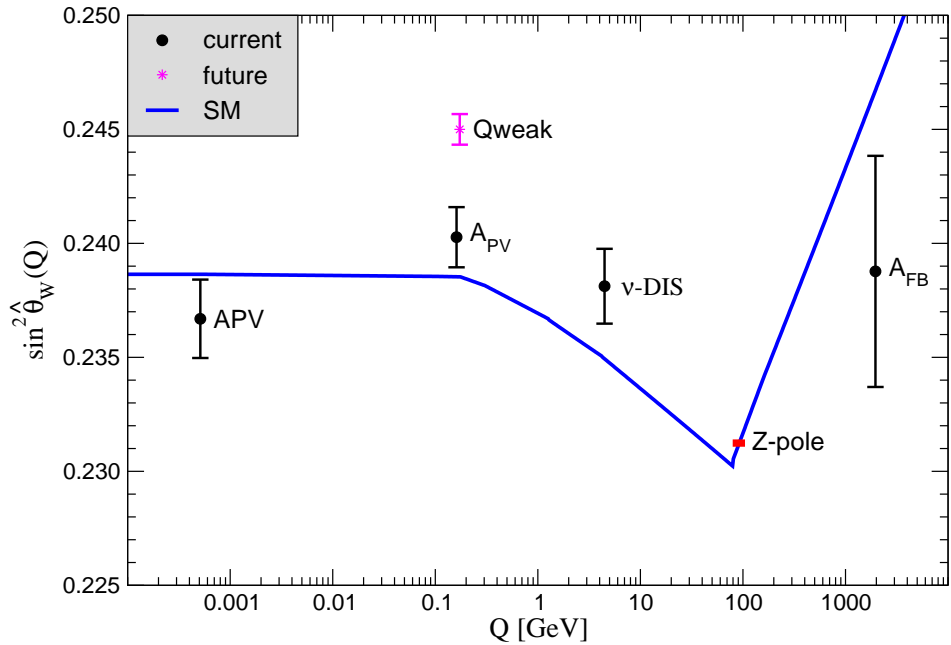}
\caption{Running of the weak angle in the \msb scheme, compared
to the theoretical expectation~\cite{s2wref}. APV, PV, and $\nu$-DIS refer to atomic parity
violation, the M\o ller asymmetry, and $\nu$ deep inelastic scattering, respectively.
From~\cite{general}.}
\label{s2w}
\end{center}
\end{figure}

  \item
 Although the $Z$-pole program has ended for the time being, there are
prospects for future programs using the Giga-$Z$ option at a  linear collider, which
might yield a factor $10^2$ more events. This would 
enormously improve the sensitivity~\cite{giga}, but would also require a large
theoretical effort to improve the radiative correction 
calculations.

\end{itemize}

\section{Fit Results}
A global fit to all data contains more information than the individual experiments,
but care must be taken with experimental and theoretical systematics and correlations.
Here we report our most recent fits for the
Particle Data Group~\cite{general}. They utilize the fully \msb program GAPP for the radiative
corrections~\cite{Erler:1999ug}. The results are generally in good agreement
with those of the LEP Electroweak Working Group~\cite{LEPEWWG}
(which uses the on-shell renormalization scheme). However, the PDG fits use
a more complete set of low energy data, which can be important for constraining
certain types of new physics.

As of November, 2007, the result of the global fit was
\bqa
           M_H &=& 77^{+28}_{-22} \mbox{ GeV}, \nonumber \\
           m_t &=& 171.1  \pm 1.9  \mbox{ GeV} \nonumber  \\
      \alpha_s &=& 0.1217 \pm 0.0017 \nonumber \\
  \hat{\alpha}(M_Z)^{-1} & = & 127.909 \pm 0.019 \nonumber \\
   \hat{s}^2_Z &=& 0.23119 \pm 0.00014 \nonumber \\
  \bar{s}^2_\ell &=& 0.23149 \pm 0.00013 \nonumber \\
          s^2_W &=& 0.22308 \pm 0.00030 \nonumber \\
 \Delta \alpha_{\rm had}^{(5)}(M_Z) &=& 0.02799 \pm 0.00014,
\label{results}
\eqa
with a good overall $\chi^2/df$ of $49.4/42$. The three values of the weak
angle $s^2$ refer respectively to the \msb, effective $Z$-lepton vertex, and on-shell
values~\cite{general}. The latter has a larger uncertainty because of a stronger
dependence on the top mass.

The precision data alone yield 
$m_t = 174.7^{+10.0}_{-7.8}$ GeV from loop corrections, in impressive
agreement with the direct Tevatron value $170.9 \pm 1.9$.
The fit actually uses the \msb \ mass $\hat m_t(\hat m_t)$, which is $\sim $ 10 GeV lower, and converts
  to the pole mass at end. The Tevatron value and the global fit value that it dominates is 
  lower than the value obtained during Run I, which leads to a lower predicted Higgs mass.

The result  \als $=0.1217 \pm 0.0017$ for the strong coupling is somewhat
above the previous world average $\als = 
0.1176(20)$, which includes other determinations, most of which
are dominated by theoretical uncertainties~\cite{hinchliffe}. 
This 
is due in part to the inclusion of the $\tau$ lifetime result~\cite{tauwidth}.
(Without it, one would obtain \als=$0.1198 \pm 0.0020$ from the $Z$-pole data.) 
The $Z$-pole value has negligible theoretical uncertainty if one
assumes the exact validity of the standard model, and is also 
insensitive to oblique (propagator) new physics. However, it is
very sensitive to non-universal new physics, such as those which affect
the $Z b \bar{b}$  vertex. The $\tau$ decay value, on the other hand, is less sensitive to new physics
but is dominated by theoretical uncertainties.

\begin{figure}[htbp]
\centering
\includegraphics*[scale=0.5]{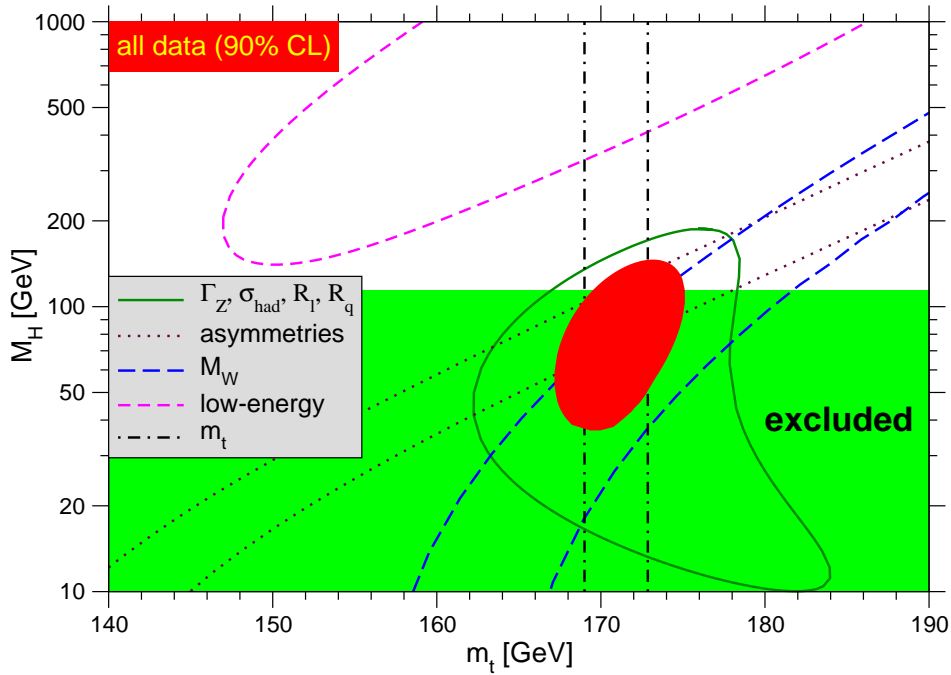}
\caption{1$\sigma$ allowed regions in \mh \  vs \mt \ and the 90\% cl global fit region from precision data,
compared with the direct exclusion limits from LEP 2, from~\cite{general}.}
\label{higgspdf}
\end{figure}

The prediction for the Higgs mass from indirect data, 
\mh $= 77^{+28}_{-22}$ GeV, should be compared with the 
direct LEP 2 limit 
$\mh \simgr 114.4\, (95\%)$ GeV~\cite{Barate:2003sz}. There is no direct conflict given the large
uncertainty in the prediction, but the central value is in the excluded region,
as can be seen in Figure \ref{higgspdf}.
 Including the direct LEP 2 exclusion results, one finds
$\mh < 167$ GeV at 95\%. 
The theoretical range in the standard model is
 115 GeV $\simle \mh \simle$ 750 GeV,
where the lower (upper) bound is from vacuum stability (triviality).
In the MSSM, one has a theoretical upper limit
 $\mh \simle 130$ GeV, while \mh \ can be as high as 150 GeV in generalizations.
 In the decoupling limit
 in which the second Higgs doublet is much heavier
 the direct search lower limit is similar to the standard model. However, the
 direct limit is
  considerably lower in the non-decoupling region in which the new supersymmetric particles and second Higgs 
  are relatively light~\cite{Heinemeyer:2007bw,Barate:2003sz}.
\mh \ enters the expressions for the radiative corrections logarithmically. It is fairly 
robust to many types of new physics, with some exceptions. In particular,
a much larger \mh \
 would be allowed for
 negative values for the $S$ parameter or positive values for $T$.
The predicted value would decrease if new physics accounted for
the value of $A_{FB}^{(0b)}$~\cite{chanowitz}.
 
%:Figure mhmt
%:Figure mwmt
\begin{figure}[htbp]
\begin{center}
\includegraphics*[scale=0.5]{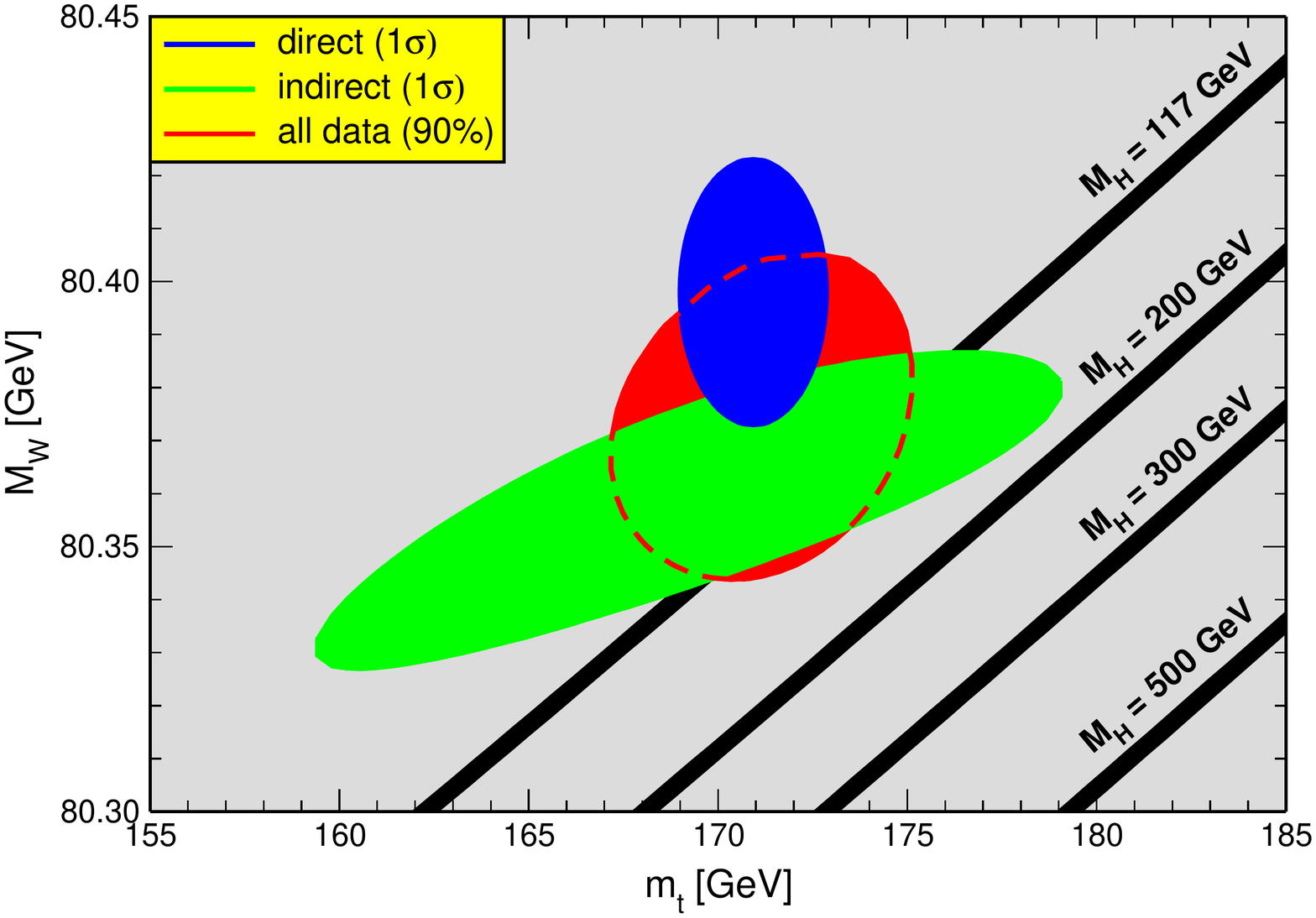}
\caption{Allowed regions in $M_W$ vs. $m_t$ from direct (Tevatron and LEP 2) and indirect data, and from the
global fit. Also shown are the standard model expectations as a function of the Higgs mass $M_H$. From~\cite{general}.}
\label{mwmt}
\end{center}
\end{figure}

\section{Beyond the Standard Model}
\begin{figure}[htbp]
\begin{center}
\includegraphics*[scale=0.5]{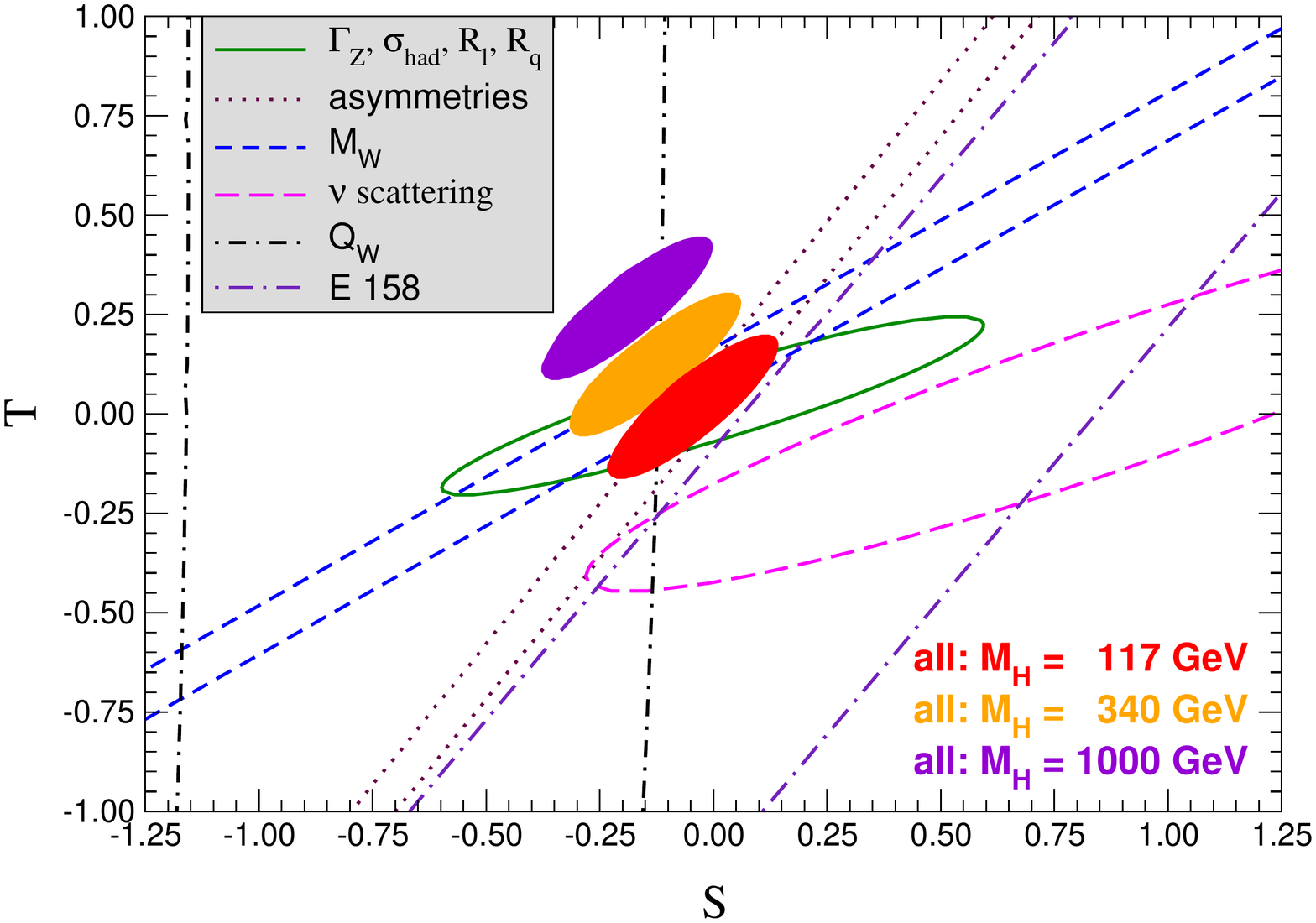}
\caption{90\% allowed contours in $S$ and $T$ from a global fit to all of the data assuming $U=0$, for $M_H=117, 340,$ 
and 1000 GeV. Also shown are the $1\sigma$ constraints from individual inputs for $M_H = 117$~GeV. $S$ and $T$  are defined  to include
only the contributions of new physics, i.e.,  $m_t$ and $M_H$ are treated separately. From~\cite{general}.}
\label{stcontour}
\end{center}
\end{figure}

The 
$\rho_0$ or $S$, $T,$ and $ U$ parameters describe the tree level effects of
Higgs triplets, or the loop effects on the $W$ and $Z$ propagators due
to such new physics as
nondegenerate fermions or scalars, or chiral families (expected,
for example,  in extended technicolor). 
The current values are\footnote{The plot of $T$ vs. $S$ produced by the LEP Electroweak Working 
Group \cite{LEPEWWG} shows larger values, $S \sim  0.07$ and $T \sim 0.13$, based on the $Z$-pole data 
and $M_W$. We almost exactly reproduce their values for the same inputs. 
The lower values reported here are due to the inclusion of the low-energy data,
such as atomic parity violation and neutrino scattering, as well as allowing 
$\alpha_s$ to float and a different evaluation of $\Delta\alpha^{(5)}(M_Z)$. 
}:
 \bqa  S &=& -0.04 \pm 0.09\ (-0.07)  \nonumber  \\
T &=& 0.02 \pm 0.09\ (+0.09)  
\eqa
for $M_H = 117$ GeV and $U=0$, where these represent the effects of new physics only (the \mt \ and \mh \ effects
are treated separately). The numbers in parentheses are the changes in the central values  when one
assumes $M_H = 300$ GeV instead.
Similarly,
 $\rho_0  \sim 1 + \alpha T =1.0004^{+0.0008}_{-0.0004}$  and 114.4 GeV $<M_H<$ 215 GeV
 (for $S=U=0$), implying limits on doublet mass splittings $\sum_i C_i \Delta m_i^2/3 < (98\ {\rm GeV})^2$ at 95\% cl, where $C_i=1(3)$ for leptons (quarks).
The standard model Higgs mass limits are weakened or can be evaded entirely for $S<0$ and $T> 0$,
as can be seen in Figure \ref{stcontour}. Most types of new physics lead to positive contributions
to $S$,  but negative values can be obtained due to Higgs doublet or triplet loops or 
Majorana fermions.
%:Figure stcontour

The $S$  constraints strongly restrict the possibility of additional chiral fermions. For example, a
 {\em degenerate} heavy family or mirror familiy is excluded at 6$\sigma$. A nondegenerate
 additional family (with a neutrino mass heavier than $M_Z/2$ to evade the lineshape constraint)
 can to some extent balance the effects of $S>0$ and $T>0$~\cite{fourth}, but is still significantly disfavored
 compared to the standard model fit~\cite{general}.

 In the 
decoupling limit of supersymmetry, in which the sparticles and second Higgs doublet are heavier than
$ \simgr 200-300$ GeV, there is little effect on the precision
observables, other than that there is necessarily 
  a light SM-like Higgs, consistent with the data. There is little
improvement on the SM fit, and in fact one can somewhat constrain
the supersymmetry breaking parameters~\cite{Heinemeyer:2007bw,susy}. However, in the
light Higgs/sparticle limit the tension between the direct Higgs searches and the
indirect precision results is relaxed. Light sparticles could also account for a muon
magnetic moment anomaly.

Heavy $Z'$ bosons are predicted by many 
grand unified  and string theories~\cite{heavyz}. Limits on the $Z'$ mass
are model dependent, but are typically  around $M_{Z'} > 800-900 $ GeV 
 from direct searches at the Tevatron, with (usually) weaker limits
from indirect constraints from WNC and  LEP~2 data. The $Z$-pole data
severely constrains the $Z-Z'$ mixing, typically
 $|\theta_{Z-Z'}| < {\rm few} \times 10^{-3}$.
A heavy $Z'$ would have many other theoretical  and
experimental implications~\cite{heavyz}.

Precision data  constrains mixings between ordinary and exotic
fermions, large extra dimensions, 
new four-fermion operators, and leptoquark bosons~\cite{general}.

Gauge unification is predicted in GUTs and some string theories.
The simplest non-supersymmetric unification is excluded by
the precision data. For the MSSM, and assuming 
no new thresholds between 1 TeV and the unification scale, one
can use the precisely known $\alpha$ and $\hat{s}^2_Z$
to predict $\als = 0.130 \pm 0.010$ and a unification scale 
$M_G \sim 3 \times 10^{16}$ GeV~\cite{polonsky}. The \als \ uncertainties are 
mainly theoretical, from the TeV and GUT thresholds, etc.
\als \ is high compared to the experimental value, but barely consistent 
given the uncertainties. 
$M_G$ is reasonable for a GUT (and
is consistent with simple seesaw models of neutrino mass),
but is somewhat below the expectations $\sim 5 \times 10^{17}$ GeV of the simplest
perturbative heterotic string models. However, this is only a
10\% effect in the appropriate variable
$\ln M_G$. The new exotic particles often present in such models
(or higher Ka\v c-Moody levels) can easily shift the $\ln M_G$
and \als \ predictions significantly, so the problem is really
why the gauge unification works so well.
It is always possible that the apparent success is accidental
(cf., the discovery of Pluto).

\section{Conclusions}

The precision $Z$-pole, LEP~2, WNC, and Tevatron experiments have
successfully tested the SM at the 0.1\% level, including electroweak loops, thus
confirming the gauge principle,
SM   group, representations, and the
basic structure of renormalizable field theory.
The standard model parameters $\sin^2 \theta_W$, $m_t$, and $\alpha_s$
were precisely determined.
In fact, \mt \ was successfully predicted from its indirect loop effects prior
to the direct discovery at the Tevatron, while the indirect value of \als,
mainly from the $Z$-lineshape, agreed with more direct QCD determinations.
Similarly, \delhad \ and $ M_H$ were constrained.
The indirect (loop) effects implied $M_H =77^{+28}_{-22}$ GeV, while direct
searches at LEP~2 yielded $M_H > 114.4 $ GeV, with a hint of a signal at 116 GeV.
The combined direct and indirect data imply $M_H < 167$ GeV at 95\% c.l.
This range is consistent with, but does not prove, 
the expectations of the supersymmetric
extension of the SM (MSSM), which predicts a light SM-like Higgs for much of
its parameter space. The agreement of the data with the SM imposes
a severe constraint on possible new physics at the TeV scale,
and points  towards decoupling theories (such as most versions of
supersymmetry and unification), which typically lead to 0.1\% effects,
rather than new TeV-scale dynamics or compositeness (e.g., Little Higgs, dynamical symmetry breaking
or composite fermions), which usually (but not always) imply  deviations of several \%, and often
large flavor changing neutral currents. 
Finally, the precisely measured gauge couplings were consistent with the
simplest form of grand unification if the SM is extended to the MSSM.

\section*{Acknowledgments}
This work was supported in part 
by DGAPA--UNAM contract PAPIIT IN115207, by the
Friends of the IAS, and by NSF grant PHT-0503584.
PL is also grateful to the conference organizers for support and to
the  Aspen Center for Physics.

%:bibliography

\end{document}